\documentclass[12pt]{article}
\usepackage{fullpage}
\usepackage{graphicx}
\newcommand{\be}{\begin{equation}}
\newcommand{\ee}{\end{equation}}

\font\mybb=msbm10 at 11pt

\def\bb#1{\hbox{\mybb#1}}

\renewcommand{\theequation}{\arabic{section}.\arabic{equation}}
\newcommand{\news}{\setcounter{equation}{0}}
\def\ben{\begin{equation}}
\def\een{\end{equation}}
\def\bea{\begin{eqnarray}}
\def\eea{\end{eqnarray}}
\begin{document}
\title{
%\begin{flushright}\ \vskip -2cm {\tiny{\em DRAFT V2}}\end{flushright}
\vskip 2cm {\bf KINKY VORTONS}\\[10pt]}
\author{
Richard A. Battye$^{1}$
and Paul M. Sutcliffe$^{2}$\\[10pt]
\\{\normalsize $^{1}$
{\sl Jodrell Bank Centre for Astrophysics,} }
\\{\normalsize {\sl University of Manchester,
 Manchester M13 9PL, U.K.}}
\\{\normalsize {\sl Email : Richard.Battye@manchester.ac.uk}}\\
\\{\normalsize $^{2}$
{\sl \normalsize Department of Mathematical Sciences,
Durham University, Durham DH1 3LE, U.K.}}\\
{\normalsize {\sl Email: p.m.sutcliffe@durham.ac.uk}}\\[10pt]}

\date{June 2008}
\maketitle
\begin{abstract}
Cosmic vortons are closed loops of superconducting cosmic strings
carrying current and charge. Despite a large number of studies the
existence and stability of cosmic vorton solutions is still an open problem.
Numerical simulations of the nonlinear field theory are difficult to
perform in (3+1)-dimensions,
due to the existence of multiple length
and time scales.
In this paper we study a (2+1)-dimensional analogue of cosmic vortons,
which we refer to as  kinky vortons, where the cosmic string is replaced
by a kink string. Many of the expected qualitative aspects of cosmic vortons
transfer to kinky vortons, with the advantage that several approximations
used in the study of cosmic vortons can be replaced by exact results.
Furthermore, the numerical study of kinky vortons requires less computational
resources than cosmic vortons, so a number of issues can be
addressed in some depth.
The radius of the kinky vorton is determined as a function of the
charge and winding number, and it is shown that the chiral limit is
a repulsive fixed point. Stability to both axial and non-axial perturbations
is demonstrated in the electric and chiral regimes, though surpringly
long lived ringing modes are observed.
Kinky vortons which are too magnetic are shown to suffer from a pinching
instability, which results in a reduction in the winding number and
can convert magnetic into electric solutions.

\end{abstract}

\newpage
\section{Introduction}\news
Cosmic strings are topological defects which may have formed during
a phase transition in the early universe (for a review see
\cite{VS}). Witten \cite{Wi1} pointed out that if the field of a
cosmic string is coupled to another complex scalar field then a
non-dissipative current can form to produce a superconducting
string. A cosmic vorton \cite{DS2} is a stable closed loop of
superconducting cosmic string which carries both current and charge.
The idea is that the current and charge on the string provide a
force to balance the string tension and prevent its collapse.

The cosmological consequences of cosmic vortons have been well
studied \cite{BCDT} and it has been proposed that they play a role
in a number of cosmological phenomena, such as galactic magnetic
fields, high energy cosmic rays, gamma ray bursts and baryogenesis.
However, the existence and stability of cosmic vortons as classical
field theory solutions is still an open problem, and convincing
numerical evidence is still lacking. Numerical simulations of the
relevant nonlinear field theory are difficult to perform in
(3+1)-dimensions and results are limited, mainly due to the
existence of multiple length and time scales. Even in the simplest
model, which is the global version of Witten's $U(1)\times U(1)$
model \cite{Wi1}, cosmic vorton solutions have not yet been
constructed. The only field theory computation to date \cite{LS} is
in a modified version of this global theory, in which the
interaction term between the two complex scalar fields is replaced
by a non-renormalizable interaction, designed to make the numerical
problem more tractable. Even in this case the vorton solution
presented eventually decays, though this may be due to
numerical issues.

In this paper we study a (2+1)-dimensional analogue of cosmic
vortons, in which the cosmic string is replaced by a kink string.
For this reason we name these objects kinky vortons. We shall show
that many of the expected qualitative features of cosmic vortons can
be demonstrated explicitly for kinky vortons, with the advantage
that several approximations used in the study of cosmic vortons can
be replaced by exact results. Furthermore, the numerical study of
kinky vortons obviously requires less computational resources than
cosmic vortons, so a number of issues can be addressed in detail. In
particular, the simulations presented in this paper reveal some of
the computational difficulties involved and help explain why similar
numerical investigations in (3+1)-dimensions have not yet been
successfully performed.

Our results include the determination of the radius of a kinky
vorton as a function of the charge and winding number, and the
demonstration that the chiral limit is a repulsive fixed point.
Stability to both axial and non-axial perturbations is exhibited in both
the electric and chiral regimes, though surpringly long lived
ringing modes are observed. Kinky vortons which are too magnetic are
found to suffer from a pinching instability, which results in a
reduction in the winding number allowing conversion of magnetic into
electric solutions. Finally, we discuss the implications of our
results on kinky vortons to possible future work on cosmic vortons.

\section{The model}\news
The Lagrangian density of our (2+1)-dimensional model is given by
\be
{\cal L}=\partial_\mu\phi \partial^\mu \phi
+\partial_\mu \sigma \partial^\mu \bar\sigma
-\frac{\lambda_\phi}{4}(\phi^2-\eta_\phi^2)^2
-\frac{\lambda_\sigma}{4}(|\sigma|^2-\eta_\sigma^2)^2
-\beta\phi^2 |\sigma|^2+\frac{\lambda_\sigma}{4}\eta_\sigma^4
\label{lag}
\ee
where $\phi$ and $\sigma$ are real and complex scalar fields respectively,
with $\eta_\phi,\eta_\sigma,\lambda_\phi,\lambda_\sigma,\beta$ all real positive
constants. This Lagrangian density
can be obtained from the global version of Witten's
$U(1)\times U(1)$ model \cite{Wi1} by restricting one of the complex scalar
fields to be real.

The theory has a global $\bb{Z}_2\times U(1)$ symmetry and the parameters
of the model can be arranged so that in the vacuum the $\bb{Z}_2$ symmetry is
broken, $\phi=\pm\eta_\phi\ne 0,$ while the $U(1)$ symmetry remains unbroken,
$|\sigma|=0.$
For this symmetry breaking pattern there exist kink strings constructed
from the $\phi$ field. If  the infinite kink string lies along the $y$-axis,
then it is given by the solution
\be
\phi=\eta_\phi\tanh\bigg(\frac{\eta_\phi\sqrt{\lambda_\phi}x}{2}\bigg), \quad
\sigma =0.
\label{barekink}
\ee
The situation of interest is when a condensate of the $\sigma$ field
carrying current and charge forms in the core of the kink string.
For the infinite string given above, such a condensate field takes the form
\be
\sigma=e^{i(\omega t+ky)}|\sigma|,
\label{condensate}
\ee
where $|\sigma|$ is a function of $x$ only with
$|\sigma|\rightarrow 0$ as $|x|\rightarrow\infty.$
The constant $k$ describes the rate of twisting of the condensate
along the string, though in the literature this is referred to as winding,
rather than twisting, so we will stick to this common convention.

A non-zero value of $\omega$ induces a charge $Q$ associated with the
global $U(1)$ symmetry, and the winding $k$ generates a current along
the string.
It is easy to see that charge and current will have opposite effects on
the string, so it is useful to introduce the combination
\be
\chi\equiv\omega^2-k^2.
\label{chi}
% \quad k=\frac{N}{R}
\ee
In the literature solutions with $\chi=0$ are termed chiral, whereas
solutions with $\chi>0$ are referred to as electric and those with
$\chi<0$ are called magnetic \cite{CP,LS}.

It is also convenient to introduce the quadratic coefficients for both fields
by the definitions
\be
m_\phi^2\equiv\frac{\lambda_\phi}{2}\eta_\phi^2, \quad
m_\sigma^2\equiv\frac{\lambda_\sigma}{2}\eta_\sigma^2+\chi.
\label{mass}
\ee
Taking into account the form of the condensate (\ref{condensate}), a
 simple analysis of the potential term in (\ref{lag}), reveals that
the condition for the $\bb{Z}_2$ symmetry to be broken in the vacuum
is
\be
\frac{m_\phi^4}{\lambda_\phi}>\frac{m_\sigma^4}{\lambda_\sigma}.
\label{cond1}
\ee
Similarly, for the $U(1)$ symmetry to remain unbroken in the vacuum
requires that the effective mass term (including a contribution from the interaction) for the $\sigma$ field is positive,
which results in the constraint
\be
\beta > \frac{\lambda_\phi m_\sigma^2}{2m_\phi^2}.
\label{cond2}
\ee
Note that both conditions (\ref{cond1}) and (\ref{cond2}) are
identical to those that arise in the cosmic vorton case \cite{LS}.

For a condensate to form in the string core we require that the solution
(\ref{barekink}) with no condensate is an unstable solution of the
field equations
\bea
\partial_\mu\partial^\mu\phi+\frac{\lambda_\phi}{2}(\phi^2-\eta_\phi^2)\phi
+\beta\phi|\sigma|^2&=&0, \label{field1}\\
\partial_\mu\partial^\mu\sigma
+\frac{\lambda_\sigma}{2}(|\sigma|^2-\eta_\sigma^2)\sigma
+\beta\phi^2\sigma&=&0.
\label{field2}
\eea
Linearizing these equations around the solution (\ref{barekink}) with
a $\sigma$ field of the form
\be
\sigma =e^{i\Omega t}e^{i(\omega t+ky)}|\sigma|,
\ee
where again $|\sigma|$ is a function of $x$ only, yields the eigenvalue
equation
\be
-|\sigma|''+(\beta\eta_\phi^2\tanh^2
\bigg(\frac{\eta_\phi\sqrt{\lambda_\phi}x}{2}\bigg)
-\frac{\lambda_\sigma}{2}\eta_\sigma^2-\chi)|\sigma|=\Omega^2|\sigma|
\ee
This is a classic example of a stationary Schr\"odinger problem
where the spectrum is known explicitly \cite{LL}. Requiring a
negative mode, $\Omega^2<0,$ gives the condition
\be
\beta < \frac{\lambda_\phi m_\sigma^2 (2 m_\sigma^2+m_\phi^2)}
{2 m_\phi^4}.
\label{cond3}
\ee
This is similar to the condition derived for cosmic vortons
\cite{HHT,LL}, but in that case some approximations must be made,
whereas here this is an exact result.

The three conditions (\ref{cond1}),(\ref{cond2}),(\ref{cond3})
provide constraints on the constants of the theory that must be satisfied
for some range of the parameter $\chi.$

\section{Exact solutions}\news
For an infinite static string, where the condensate field has the form
(\ref{condensate}), the field equations (\ref{field1}) and (\ref{field2})
reduce to the following nonlinear coupled ordinary differential equations
\bea
\phi''&=&\phi\bigg(\frac{\lambda_\phi}{2}(\phi^2-\eta_\phi^2)
+\beta|\sigma|^2\bigg)
\label{ode1}\\
|\sigma|''&=&|\sigma|\bigg( -\chi
+\frac{\lambda_\sigma}{2}(|\sigma|^2-\eta_\sigma^2)+\beta\phi^2
\bigg). \label{ode2} \eea 
If $\chi=0$ then an exact solution to the above equations can be found if \cite{Ra,Ho}
\be
\left(2-{\lambda_\sigma\over\beta}\right)\eta_{\sigma}^2=\left({\lambda_\phi\over 2\beta}
+{2\beta\over\lambda_\sigma}-2\right)\eta_\phi^2\,.
\ee
For the purposes of the current investigation
it is useful if exact solutions are available for a range of values
of the parameter $\chi.$ This can only be achieved for the choice \be
2\beta=\lambda_\phi=\lambda_\sigma\equiv\lambda. \label{paramx} \ee
With this relation between the constants of the theory a
simplification occurs and the two conditions (\ref{cond1}) and
(\ref{cond2}) become identical and reduce to the condition $
m_\phi^2 > m_\sigma^2.$ Furthermore, the final condition
(\ref{cond3}) simplifies to $m_\phi^2 < 2 m_\sigma^2,$ so the
whole set of conditions reduces simply to the constraint \be
\frac{1}{\sqrt{2}}m_\phi  < m_\sigma < m_\phi. \label{cond5} \ee The
exact solutions to (\ref{ode1}) and (\ref{ode2}) are given by \be
\phi=\eta_\phi\tanh\bigg(x\sqrt{m_\phi^2-m_\sigma^2}\bigg),
\quad\quad
|\sigma|=\sqrt{\frac{2(2m_\sigma^2-m_\phi^2)}{\lambda}}{\rm sech}\,
\bigg(x\sqrt{m_\phi^2-m_\sigma^2}\bigg), \label{gstring} \ee which
are valid for the entire range of $\chi$ satisfying the condition
(\ref{cond5}).

At this stage there are three remaining constants of the theory, that is,
$\eta_\phi,\eta_\sigma,\lambda.$
Most of the analysis which follows can be extended to arbitrary values
of these constants (providing they satisfy (\ref{cond5}) for some range
of $\chi$) but in order to simplify the presentation and to compare with
later numerical simulations we shall now fix some specific values.
In the penultimate section we shall discuss how other choices for the
constants of the theory influence the results.

In the chiral case, $\chi=0,$ the condition (\ref{cond5}) reduces to
\be
\frac{1}{2} < \bigg(\frac{\eta_\sigma}{\eta_\phi}\bigg)^2 <1.
\label{cond6}
\ee
To aim for the most generic behaviour possible we set
$\eta_\sigma=\sqrt{3}\eta_\phi/2,$ so that the value chosen for the
ratio which appears in (\ref{cond6}) is at the midpoint of the allowed range.
Finally, the overall scales are normalized by setting
$\eta_\phi=1$ and $\lambda=2.$
In summary, the constants selected in the remainder of this paper (results for more general choices are presented in appendix A) are
\be
\eta_\phi=1, \quad \eta_\sigma=\sqrt{3}/2, \quad \lambda_\phi=2, \quad
 \lambda_\sigma=2,\quad \beta=1.
\label{param1}
\ee
With the above choice the constraint (\ref{cond5}) reduces to
\be
-\frac{1}{4}<\chi<\frac{1}{4}
\label{limits}
\ee
hence a reasonable range of electric and magnetic solutions are
possible in addition to the chiral case.
The infinite string solutions (\ref{gstring}) simplify to
\be
\phi=\tanh\bigg(\frac{x\sqrt{1-4\chi}}{2}\bigg), \quad\quad
|\sigma|=\sqrt{\frac{1+4\chi}{2}}{\rm sech}\,
\bigg(\frac{x\sqrt{1-4\chi}}{2}\bigg).
\label{string}
\ee
From these exact solutions the result of adding current ($\chi<0$) and
charge ($\chi>0$) to the condensate can easily be seen.
Current clearly quenches the condensate, with the maximum amplitude decreasing
as $\chi$ decreases, until finally the condensate disappears in the limit
as $\chi\rightarrow -1/4,$ which is the lower edge of the allowed range
(\ref{limits}). There is also a moderate decrease of the width of the
condensate (and kink) as the current is increased.
On the other hand, charge enhances the condensate,
with the maximum amplitude increasing as $\chi$ increases.
Moreover, the width of the condensate (and kink) increases dramatically
as $\chi$ increases, and becomes delocalized in the limit
$\chi\rightarrow 1/4,$  which is the upper edge of the allowed range
(\ref{limits}). Similar quenching and anti-quenching effects are seen
in the numerical solutions of infinite cosmic strings carrying a condensate
in the (3+1)-dimensional model \cite{LS}.

\section{Kinky vortons}\news
We now turn our attention to kinky vortons, and apply some approximations,
in conjunction with the exact results of the previous section, to study
circular loops of kink strings carrying a condensate. In the subsequent discussion we will use the terms vortons and cosmic vortons to refer to kinky vortons and their three-dimensional analog respectively.

In polar coordinates the ansatz for a circular loop reads
\be
\phi=\phi(r), \quad \sigma=e^{i(\omega t+N\theta)}|\sigma|(r),
\label{radsym}
\ee
with the boundary conditions $\phi'(0)=0,\ \phi(\infty)=1$ and
$|\sigma|(0)=0,\ |\sigma|(\infty)=0.$ This describes a vorton of
radius $R$ where $\phi(R)=0.$ The integer $N$ is referred to as the
winding number. Note that to compare with the straight string the
winding rate $k$ depends upon both $N$ and the loop radius $R$ since
$k={N}/{R}.$

The conserved charge $Q$ of such a field is given by
\be
Q=\frac{1}{2i}\int (\dot\sigma\bar\sigma-\dot{\bar\sigma}\sigma)\, d^2x
=2\pi\omega \int_0^\infty |\sigma|^2r\,dr.
\label{charge1}
\ee
The main aim is to determine the properties of a vorton, and in particular
its radius $R,$ as a function of the conserved quantities $Q$ and $N.$

If the radius $R$ is much larger than the width
of the condensate (and kink) then a section through the circular loop
can be well-approximated by a section through the infinite straight string.
This means that approximations of the following form can be used
to evaluate a number of required integrals
\be
 \int_0^\infty |\sigma|^p r\,dr
\approx
R\Sigma_p
\equiv
R\int_{-\infty}^{\infty}
\bigg[\sqrt{\frac{1+4\chi}{2}}{\rm sech}\,
\bigg(\frac{x\sqrt{1-4\chi}}{2}\bigg)\bigg]^p\, dx,
\ee
where the infinite string solution (\ref{string}) has been used
to determine the cross-sectional contribution.
In particular, we shall later require the exact results
\be
\Sigma_2=\frac{2(1+4\chi)}{\sqrt{1-4\chi}}, \quad\quad
\Sigma_4=\frac{2(1+4\chi)^2}{3\sqrt{1-4\chi}}.
\label{sigma24}
\ee
In the analysis of cosmic vortons analogous quantities to $\Sigma_2$
and $\Sigma_4$ are also required, but in that case they can only be
computed numerically; although analytical estimates have been given
\cite{LS} which have a similar form to the exact results (\ref{sigma24}).
The main qualitative difference is the appearance of the square root
in the denominators in (\ref{sigma24}), which can be attributed to the reduction in the number of spatial dimensions.

Substituting the above result for $\Sigma_2$ into (\ref{charge1}) yields
\be
Q=4\pi\omega R\frac{(1+4\chi)}{\sqrt{1-4\chi}}.
\label{charge2}
\ee
In the chiral case, $\chi=0,$ then $\omega=k=N/R,$ thus (\ref{charge2})
reveals that for chiral vortons the relationship between the
conserved quantities is $Q=4\pi N,$ and is independent of $R.$
Thus loops which are initially chiral remain chiral even if the radius
changes.

The first goal is to determine the radius $R$ at which the vorton energy
is minimized for fixed values of the conserved quantities $Q$ and $N.$
One difficulty is that the cross-sectional properties of the string
(for example, the amplitude and width of the condensate) vary with the vorton
radius. As seen above, this variation is most easily quantified as a
function of $\chi,$ and is highly implicit as a function of $Q$ and $N.$
This is a serious obstacle for cosmic vortons, but the fact that we have
exact expressions available makes the problem much more tractable.

Taking the square of equation (\ref{charge2}) and making the substitution
$\omega^2=\chi+N^2/R^2$ generates the cubic
\be
16\chi^3+8\bigg(1+\frac{2N^2}{R^2}\bigg)\chi^2
+\bigg(1+\frac{8N^2}{R^2}+\frac{Q^2}{4\pi^2R^2}\bigg)\chi
+\bigg(N^2-\frac{Q^2}{16\pi^2}\bigg)\frac{1}{R^2}=0.
\label{cubic}\ee
Given the physical properties of the vorton, that is the conserved quantities
$Q$ and $N$ together with the radius $R,$ this cubic determines $\chi$
(as the root which lies in the interval $(-1/4,1/4)$).

\begin{figure}
\begin{center}
\includegraphics[width=12cm]{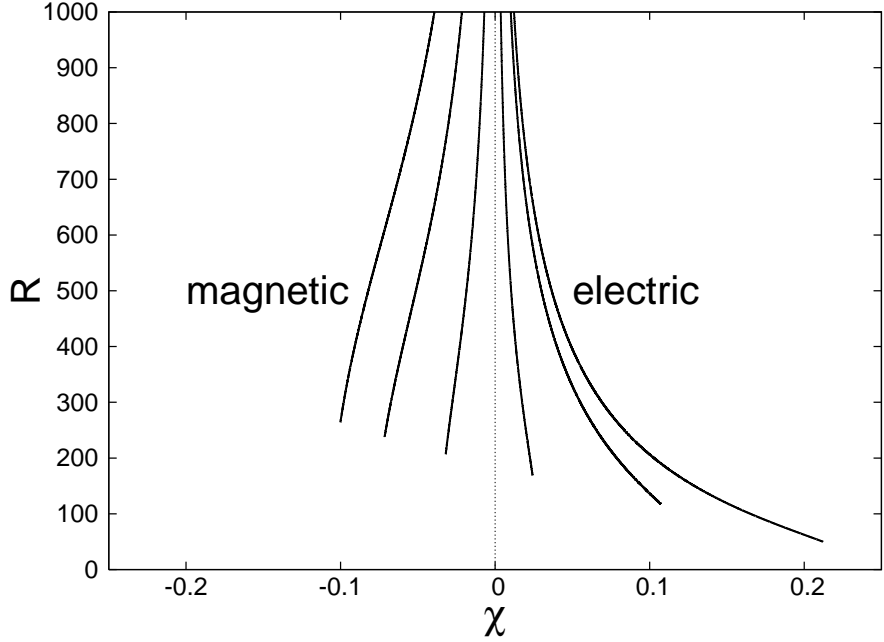}
\caption{The variation of $\chi$ as the vorton loop radius $R$ decreases
for $Q=1500$ and $N=250,200,150,100,50,10$ (left to right).
Note that as $R$ decreases each curve moves away from the chiral
line $\chi=0.$ Each curve terminates at the energy minimizing radius.
}
\label{fig-repel}
\end{center}
\end{figure}

The cubic (\ref{cubic}) allows the variation of $\chi$ with the radius
$R$ to be easily studied. First of all, the cubic confirms that the chiral
case is a fixed point, that is, if $Q=4\pi N$ then $\chi=0$ for all $R.$
Also, as $R\rightarrow\infty$ the final term in (\ref{cubic}) tends
to zero, thus $\chi\rightarrow 0,$ demonstrating that if a vorton is
created with a sufficiently large radius it will be arbitrarily close
to chiral. Early studies of cosmic vortons suggested that the chiral limit
might be an attractor \cite{DS2} but more recent investigations \cite{LS}
have demonstrated that the chiral limit is a repulsive fixed point not
an attractor. For kinky vortons
this can easily be seen from (\ref{cubic}) as follows.
The electric regime is given by $Q>4\pi N,$ in which case the final
term in (\ref{cubic}) is negative and $\chi>0.$ For large enough $R$ then
$\chi\approx 0,$ but as $R$ decreases the modulus of the final term in
(\ref{cubic}) increases and therefore so does $\chi,$ moving away from
the chiral limit. Similarly, if  $Q<4\pi N,$ then all coefficients in
the cubic (\ref{cubic}) are positive thus $\chi<0$ and this is the
magnetic regime. As $R$ decreases then again the modulus of the
final term in (\ref{cubic}) increases and therefore $\chi$ becomes
more negative, again moving away from
the chiral limit.
This is illustrated graphically in Figure~\ref{fig-repel}, where
curves are plotted showing the value of $\chi$ as $R$ decreases for
$Q=1500$ and $N=250,200,150,100,50,10$ (left to right).
In each case as $R$ decreases the curve moves away from the chiral
line $\chi=0.$ Each of these curves terminates at the energy minimizing
radius, which we now discuss.

Using the cross-sectional formulae (\ref{sigma24}) and similar
expressions for the cross-sectional integrals of powers of $\phi,\phi'$
and $|\sigma|'$ the vorton energy is found to be
\be
E=\frac{4\pi}{\sqrt{1-4\chi}}\bigg(
\frac{R}{3}\bigg[\frac{5}{2}-5\chi+4\chi^2\bigg]
+\frac{1}{R}\bigg[N^2(1+4\chi)
+\bigg(\frac{Q}{4\pi}\bigg)^2\frac{(1-4\chi)}{(1+4\chi)}
\bigg]
\bigg).
\label{aenergy}
\ee
\begin{figure}
\begin{center}
\includegraphics[width=12cm]{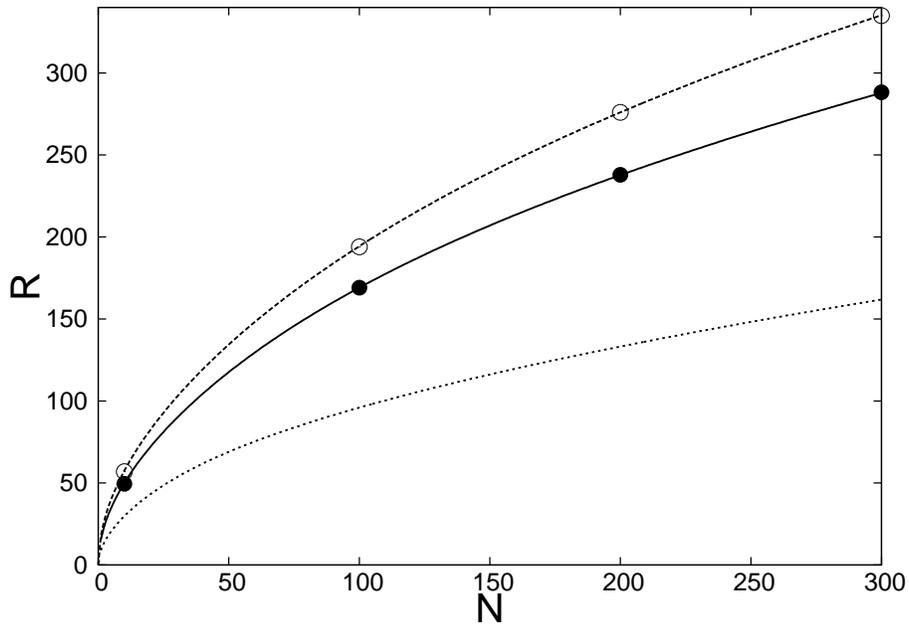}
\caption{
The energy minimizing radius $R$ as a function of $N$ for the
three values $Q=2000$ (dashed curve), $Q=1500$ (solid curve), and
$Q=500$ (dotted curve). The results of a field theory energy minimization
are presented as filled circles for $Q=1500$ and open circles for
$Q=2000.$ 
We note that the solution with $Q=1500,$ 
$N=10$ and $R=50$ is the smallest radius 
solution we have been able to find. 
Vorton solutions will only exist for points on each of these curves
down to a minimal value of $R$ due to the splitting instability.
}
\label{fig-RQ}
\end{center}
\end{figure}
For given values of the conserved quantities $Q$ and $N,$ this defines
the energy as a function of the vorton radius $R,$ where $\chi$ is
again given by (\ref{cubic}). Using this formula the value of $R$
at which this energy is minimized can be calculated.
This is plotted in Figure~\ref{fig-RQ} for $1\le N\le 300$ and
the three values $Q=2000$ (dashed curve), $Q=1500$ (solid curve) and
$Q=500$ (dotted curve).

The energy formula (\ref{aenergy}) together with the cubic (\ref{cubic})
can be used to show that if either $Q=0$ or $N=0$ then the energy minimizing
radius is $R=0.$ Thus vortons require both current and charge, in agreement
with the results for cosmic vortons \cite{LS}.

The above analysis for the vorton radius as a function of $Q$ and $N$
can be compared with the results of numerical simulations of the full nonlinear
field theory. Restricting to radially symmetric fields of the form
(\ref{radsym}) the energy can be written as
\be
E=2\pi\int_0^\infty\left\{
\phi'^2+|\sigma|'^2+\frac{1}{2}(\phi^2-1)^2
+|\sigma|^2\left(\phi^2+\frac{N^2}{r^2}+\frac{1}{2}|\sigma|^2-\frac{3}{4}\right)\right\}
r\,dr +\frac{Q^2}{2\pi\int_0^\infty |\sigma|^2r\,dr}.
\label{radenergy}
\ee
For fixed values of $N$ and $Q$ this energy has been minimized using a
simulated annealing algorithm. The resulting loop radius for 
the four values $N=10,100,200,300$ are presented in Figure~\ref{fig-RQ}
as the filled circles for $Q=1500$ and the open circles for $Q=2000.$
  From these, and other similar results, it can be seen
that the analytic calculation compares well with the numerical results,
when a vorton solution exists. However, if $Q$ and $N$ are not sufficiently
large then it appears that no vorton solution exists.
For example, for $Q=500$ (corresponding
to the dotted curve in Figure~\ref{fig-RQ}) no solutions were found for
any value of $N$ and for $Q=1500$ no solution was found with $N<10$. 
This is because of a splitting instability where
the effective potential trapping the condensate on the kink loop is
not strong enough to prevent a separation in which the
condensate radiates away to infinity leaving behind a bare kink loop that
collapses. This instability can not be seen in the analytic calculation,
 which assumes that the condensate is trapped on the kink loop, but is clearly
evident in the numerical energy minimization process. In principle, it
should be possible to numerically map out the region of the $(Q,N)$-plane
in which vortons exist. We have not performed this computation due
to the fact that it requires significant calculation to determine the boundary
of instability since it can take a long time to develop when close to the critical value.

\section{Field theory dynamics}\news
In this section we describe the results of numerical simulations of
the dynamical field theory equations (\ref{field1}) and
(\ref{field2}). The numerical algorithm is an explicit finite
difference scheme which is second order accurate in both the space
and time derivatives. For most simulations the lattice spacing is
taken to be $\Delta x=0.5,$ although for some simulations this is
reduced to $\Delta x=0.25$ when greater resolution is required. The
time step is $\Delta t=0.1$ for all the simulations presented.
Neumann boundary conditions are applied and the number of grid
points ranges from $401\times 401$ to $2801\times 2801$ depending
upon the size of the vorton that is being simulated.

To study the stability of a vorton to axially symmetric perturbations
we wish to consider its evolution
from an initial state in which the radius differs from that of the optimal
value. Of course, the initial field must have the correct charge
$Q$ and winding $N$ of the vorton we wish to study.
To create such an initial condition for a circular loop with an initial
radius $R_0,$ we use the radially symmetric ansatz (\ref{radsym})
with the required value of $N$ and where the fields through a section of
the loop are approximated by the infinite straight string fields (\ref{string}).
Let $\chi_0$ be the value of $\chi$ used in the initial conditions, which
is not that of the original vorton to be investigated,
but is obtained by substituting the required values of $Q$ and $N,$
together with the initial radius $R_0,$
into the cubic (\ref{cubic}). The initial frequency $\omega_0$ is then obtained
as $\omega_0=\sqrt{\chi_0+(N/R_0)^2}$, where we have made use of the
formula (\ref{chi}).

\begin{figure}
\begin{center}
\includegraphics[width=12cm]{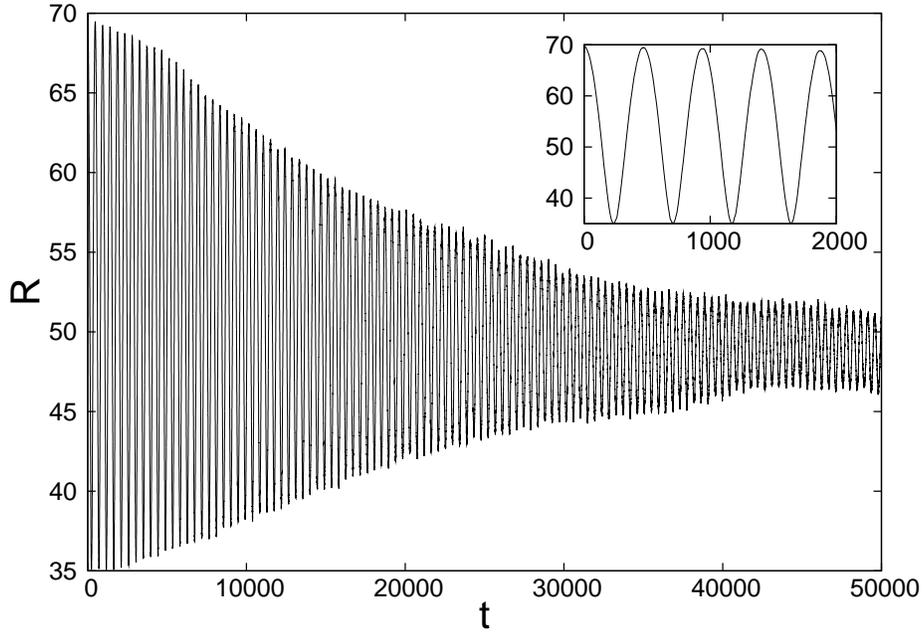}
\caption{ The radius of an electric vorton for $0\le t \le 50000.$
The charge and winding are $Q=1500$ and $N=10,$ corresponding to an
optimal radius of $R=50,$ but the initial condition has $R=R_0=70.$
The inset is a blow-up of this plot for early times $0\le t \le
2000.$ It can be seen that the radius oscillates around the optimal
value but the amplitude decreases extremely slowly. }
\label{fig-inset1}
\end{center}
\end{figure}

\begin{figure}
\begin{center}
\includegraphics[width=16cm]{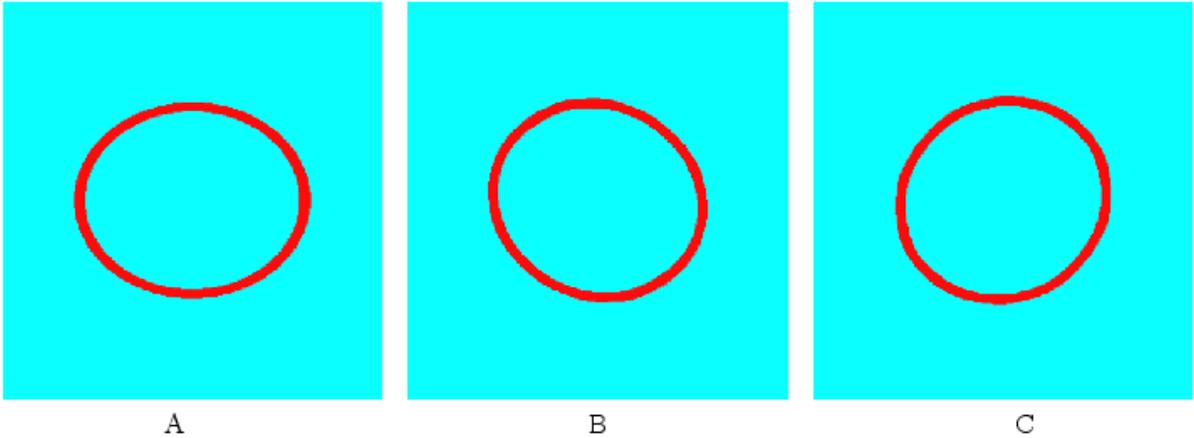}
\caption{An electric vorton at $t=0,8000,20000.$ Dark (red) regions
are where $\phi^2\le 0.2.$ The charge and winding number are
$Q=1500$ and $N=10,$ corresponding to an optimal radius of $R=50.$
The initial condition is a circular loop of radius $R=50$ which has
been squashed by $10\%$ along the $y$-axis and stretched by the same
factor along the $x$-axis.
 }
\label{fig-elec2}
\end{center}
\end{figure}

Vortons with a small loop radius require less computational resources
to study than larger vortons, simply because fewer grid points are required
to fit the vorton inside the simulation region.
From Figure~\ref{fig-RQ}, and recalling the earlier comments that
large values of $Q$ appear to be necessary to avoid the splitting instability,
it can be seen that the strongly electric regime is the easiest to study.
In particular, there is a solution with $Q=1500$ and $N=10,$ with
a corresponding value of $\chi=0.193$ and a radius $R=50,$ which is
the smallest vorton that appears to exist in this theory.
Field theory simulations of this vorton using the correct initial radius
$R_0=50$ confirm that the vorton radius remains constant throughout
the simulation. To test stability we consider an initial condition in
which the radius is $R_0=70$ and therefore is substantially larger than
the correct value for the given charge and winding. The evolution
of the radius with time is displayed in Figure~\ref{fig-inset1}.
The inset shows the evolution of the radius for earlier times $0 \le t\le 2000,$
when only a few oscillations of the radius have taken place.
This confirms that the radius oscillates around the optimal value $R=50$
but reveals that the amplitude of the oscillation decreases extremely
slowly due to radiation.
The full plot extends to very large times $0\le t\le 50000$
and reveals the decaying amplitude and approach to the optimal radius,
but even after over a hundred oscillations the amplitude is still a reasonable
fraction of the initial perturbation. The ringing modes are surprisingly long
lived and the simulations require a large number of time steps
(in this example half a million) to observe the approach to equilibrium.
\begin{figure}
\begin{center}
\includegraphics[width=11cm]{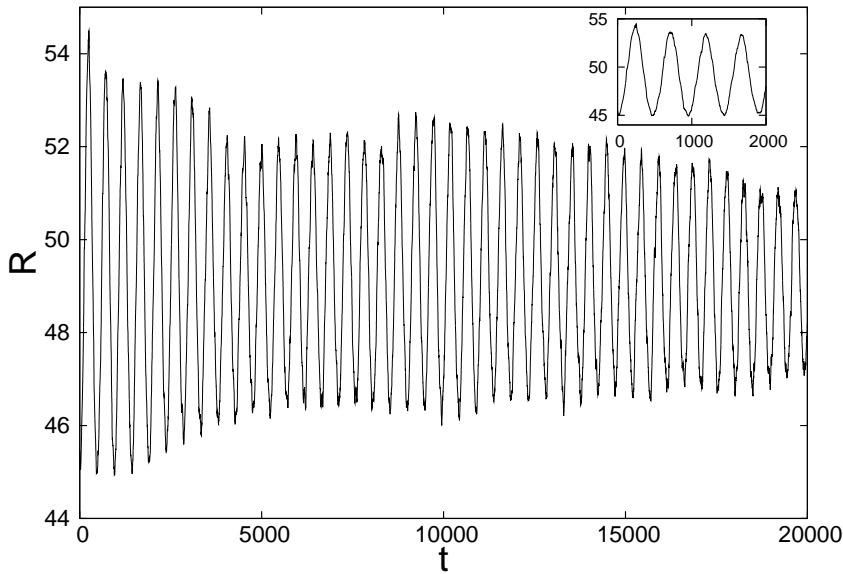}
\caption{The radius along the positive $x$-axis of an electric
vorton for $0\le t \le 20000.$ The charge and winding are $Q=1500$
and $N=10,$ corresponding to an optimal radius of $R=50,$ but the
initial condition is an ellipse. The inset is a blow-up of this plot
for early times $0\le t \le 2000.$ Long-lived ringing modes are observed once
again. } \label{fig-inset2}
\end{center}
\end{figure}

We have performed a number of simulations similar to the one
described above, and all suggest that when vorton solutions exist
they are stable to axially symmetric perturbations. In order to test
stability to general perturbations we require an initial condition
which breaks the axial symmetry, but preserves the charge
$Q$ of the unpertubed solution. Such an initial condition is
obtained by taking a vorton solution and performing the combined
scaling transformation $x\mapsto x\mu$ and $y\mapsto y/\mu,$ which
clearly leaves the charge $Q$ invariant and converts the circular
vorton into an ellipse. In the study of cosmic vortons with a
modified interaction term, elliptic initial conditions were also
considered, but in that case it appears that these were required to
be fine-tuned to ensure an initially homogeneous current
distribution. As we shall see, our vortons are more robust and do
not require any fine-tuning of the initial elliptic loop.

As an illustration, we again consider the electric vorton with
parameters $Q=1500$ and $N=10$ yielding the radius $R=50.$ The
initial condition has squashing parameter $\mu=1.1$ to produce the
ellipse displayed in Figure~\ref{fig-elec2}A. Dark regions are
where $\phi^2\le 0.2,$ and hence indicate the position of the kink
string. The vorton evolves as a rotating and oscillating ellipse,
with the initial elliptical shape still apparent even at the much
later time $t=20000$ Figure~\ref{fig-elec2}C. It is expected that
the eccentricity of the initial loop decreases with time so that the
unperturbed circular vortex is eventually recovered. However, as
with the axially symmetric perturbation, there are extremely
long-lived ringing modes. In Figure~\ref{fig-inset2} we plot the
radius along the positive $x$-axis as a function of time, to
demonstrate the slow approach towards equilibrium.

\begin{figure}
\begin{center}
\includegraphics[width=16cm]{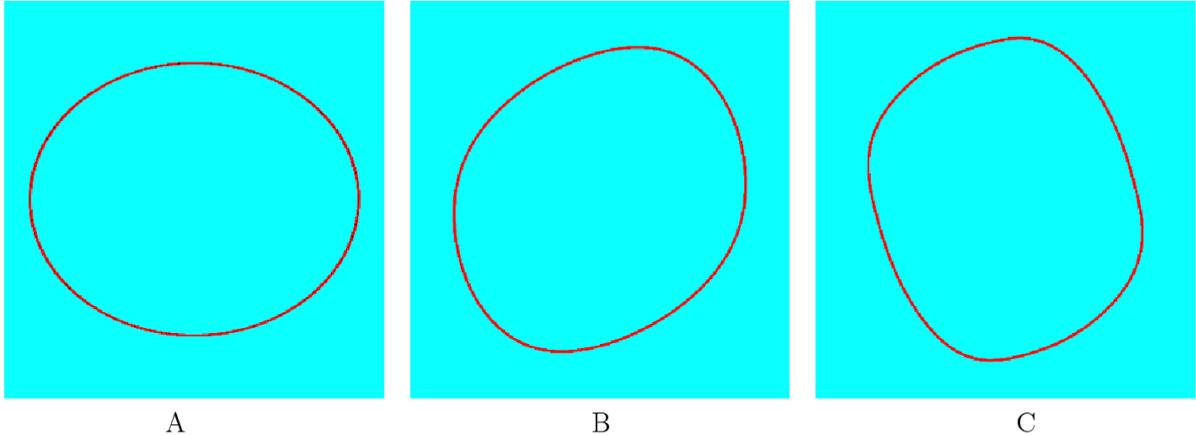}
\caption{A chiral vorton at $t=0,5000,10000.$ Dark (red) regions are
where $\phi^2\le 0.5.$ The charge and winding number are $Q=1500$
and $N=119,$ corresponding to an optimal radius of $R=185.$ The
initial condition is a circular loop of radius $R=185$ which has
been squashed by $10\%$ along the $y$-axis and stretched by the same
factor along the $x$-axis. } \label{fig-chiral1}
\end{center}
\end{figure}

The field theory dynamics described so far has been concerned with
strongly electric vortons, but similar results have also been
obtained for chiral vortons. As an example, Figure~\ref{fig-chiral1}
displays the evolution ($t=0,5000,10000$) of a squashed chiral
vorton. The unperturbed vorton has $Q=1500$ and $N=119$ giving a
radius $R=185$ and $\chi\approx 0.$ The elliptic perturbation is
given by a squashing parameter $\mu=1.1.$ The chiral vorton is
substantially larger than the strongly electric vorton studied above
with the same charge $Q.$ Recall that as $\chi$ decreases the kink
and condensate width both decrease, so that the ratio of the vorton
radius to kink width is significantly larger in the chiral case than
in the strongly electric regime. Chiral vortons are certainly well
inside the thin ring limit. This is illustrated in
Figure~\ref{fig-chiral1} where the dark region is where
$\phi^2\le 0.5.$ Note that this value of $0.5$ is larger than the
value $0.2$ used in the plots of the electric vorton, and therefore
makes the chiral vorton appear thicker than it would have done if
the same value as earlier had been used. However, for the chiral
vorton a much bigger grid is required and the results in
Figure~\ref{fig-chiral1} correspond to a grid containing
$1001\times1001$ grid points, but with the same lattice spacing
$\Delta x=0.5.$ These results confirm the stability of chiral vortons
and also demonstrate that significant computing resources are
required even to study kinky vortons in the chiral regime and that
similar studies of cosmic vortons would be prohibitive.
This helps to explain why numerical studies of
cosmic vortons are currently limited and the only
solutions found to date are in a theory with a modified interaction
term designed to reduce these difficulties. We shall make
some further remarks on this topic later.

Analytic and numerical work \cite{LS} on infinite cosmic strings
carrying current and charge has shown that if the string is strongly
magnetic $(\chi\ll 0$) then it will develop a pinching instability, in which the
condensate field is driven to zero at some point on the string and
the field partially unwinds to produce a string with a lower winding
number $N.$ Based on this result it has been conjectured that a magnetic vorton with $|\chi|>\chi_{\rm c}$  will excite the unstable pinching mode,
reduce its winding number and recover as a vorton which is less
magnetic, or possibly even electric. We shall now investigate the
dynamics of magnetic kinky vortons and demonstrate
this process has a significant impact on the dynamics.

An analysis of infinite kink strings to a pinching instability is
presented in appendix B and suggests that such an instability will
be present if $\chi<-0.1.$ This should provide a reasonable estimate
for the value at which such an instability emerges for kinky
vortons, but is not expected to be exact as it neglects the
curvature of the vorton loop. In fact the numerical simulations
described below reveal an instability when $\chi\approx-0.08,$ which is in
reasonable agreement with the analytic calculation.

A magnetic vorton will have a larger radius and a smaller width than
its chiral counterpart so even larger grid sizes must be used to
simulate the magnetic regime. The following simulations of magnetic
vortons used a grid containing $2801\times 2801$ points and a
reduced lattice spacing of $\Delta x=0.25,$ with 80000 time steps
per simulation. Once again it is worth pointing out that analogous
simulations in (3+1)-dimensions would require very considerable
computing resources.

\begin{figure}
\begin{center}
\includegraphics[width=12cm]{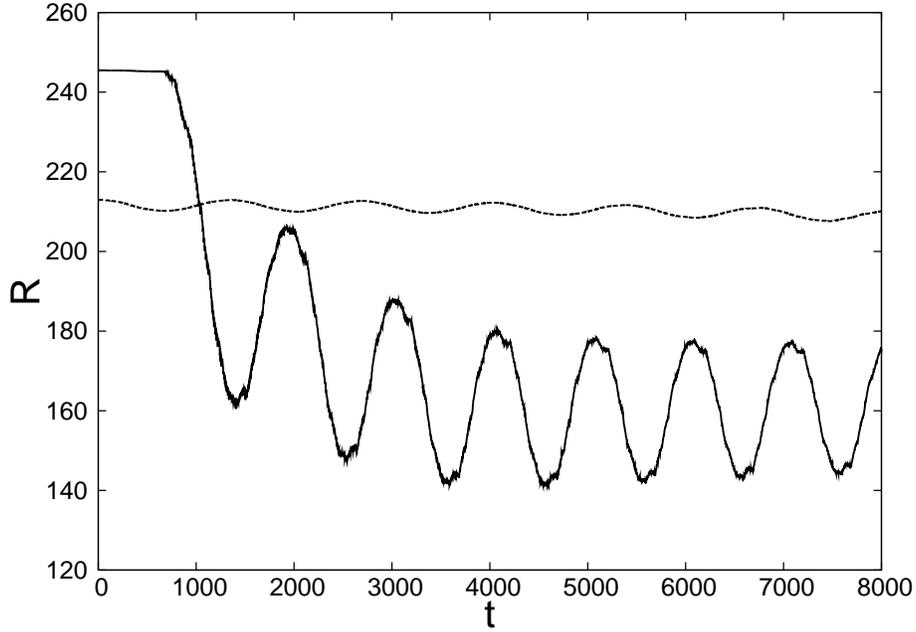}
\caption{The radius as a function of time for two initially magnetic
vortons. The dashed curve corresponds to a vorton with initial
parameters $Q=1500$ and $N=159$ giving $R=213$ and $\chi=-0.04.$
This vorton is stable and the radius oscillates around the initial
value with a small amplitude. The solid curve is for a vorton with
initial parameters $Q=1500$ and $N=214$ giving $R=245$ and
$\chi=-0.08.$ This vorton is unstable and the radius decreases
significantly near the start of the evolution and then oscillates
around the lower value of $R=167.$
 } \label{fig-mag}
\end{center}
\end{figure}

 First of all we shall consider a magnetic vorton
which is expected to be stable. If $Q=1500$ and $N=159$ then $R=213$
and $\chi=-0.04,$ which should be within the stable regime. The
dashed curve in Figure~\ref{fig-mag} shows that the radius
oscillates with a small amplitude around the initial value,
suggesting that this magnetic vorton is indeed stable.

\begin{figure}
\begin{center}
\includegraphics[width=16cm]{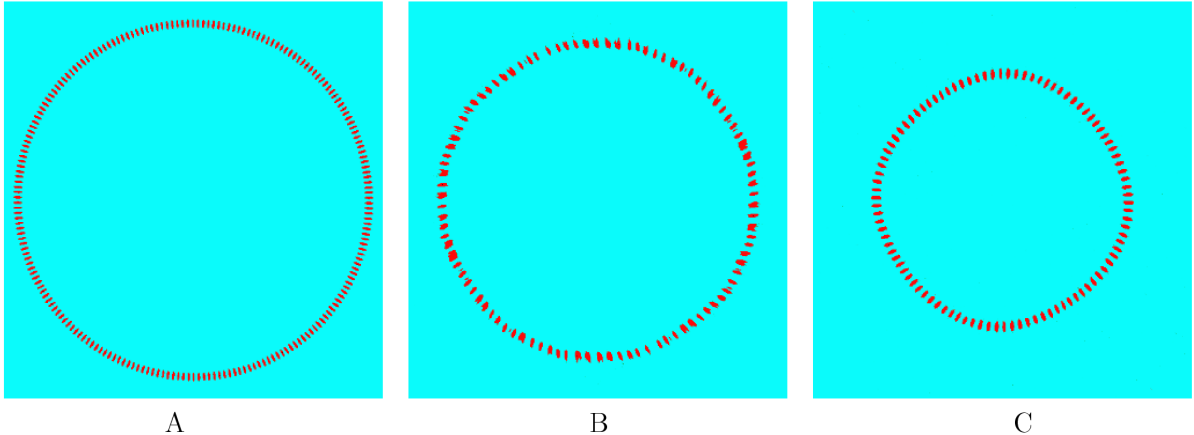}
\caption{The dark (red) region is where $\Re(\sigma)>0.05,$
plotted at the times $t=0,1000,8000.$ Initially it can be seen that
the winding number is $N=214$ but this has reduced to $N=98$ in the
two subsequent plots.} \label{fig-untwist}
\end{center}
\end{figure}

For the values $Q=1500$ and $N=214$ the vorton radius is $R=245$ and
$\chi=-0.08,$ which is expected to be close to the onset of the
pinching instability. The solid curve in Figure~\ref{fig-mag}
displays the radius as a function of time for this vorton. It can be
seen that the radius decreases significantly towards the start of
the evolution and subsequently oscillates around a much lower value.
This is precisely the behaviour expected of an unstable magnetic
vorton which partially unwinds and recovers to a vorton with a lower
value of $N.$ This interpretation is confirmed in
Figure~\ref{fig-untwist} where we plot the dark region where
$\Re(\sigma)>0.05$  at times $t=0,1000,8000.$ The advantage of
displaying the quantity $\Re(\sigma)$ is that not only can the
location of the condensate (and hence the vorton radius) be observed
but so can the winding number $N,$ since it corresponds to the
number of dashes contained in the vorton loop. Counting the number
of dashes in Figure~\ref{fig-untwist}A confirms that initially
$N=214.$ Counting the number of dashes in Figure~\ref{fig-untwist}B
and Figure~\ref{fig-untwist}C reveals that each configuration has a
winding number $N=98.$ A vorton with $Q=1500$ and $N=98$ has a
radius $R=167$ and this is consistent with the radius around which
the solid curve in Figure~\ref{fig-mag} oscillates at later times.
This vorton has $\chi=0.03$ and is therefore electric. These results
confirm that the initial magnetic vorton is unstable and
partially unwinds to convert to a stable electric vorton. Note that
the pinching instability can only be excited if the axial symmetry
of the vorton is broken but this is provided by the small
perturbation produced by the numerical grid and in particular the boundary. 
This can be seen by the square deformation
 of the vorton in Figure~\ref{fig-untwist}C.

\section{Parameter values and electric instability}\news
The field theory simulations discussed in the previous section illustrate
that the computational resources required to study strongly electric
vortons are not as substantial as for chiral or magnetic vortons.
This is relevant for the study of cosmic vortons, where simulations
in (3+1)-dimensions are close to the limit of current feasibility.
It is more likely that cosmic vortons can be constructed numerically
if they exist in the strongly electric regime. In this section we shall
discuss how the existence of strongly electric kinky vortons can depend
crucially on the specific parameter values of the model.

Recall from earlier, that vortons do not exist if they are too electric
because of a splitting instability, where
the effective potential trapping the condensate on the kink loop is
not strong enough to prevent a separation in which the
condensate radiates away to infinity. The only current numerical
construction \cite{LS}  of cosmic vortons is in a modified model where
the interaction term $|\phi|^2|\sigma|^2$ is replaced by a non-renormalizable
interaction term $|\phi|^6|\sigma|^2.$ The effect of this modification
is to increase the width of the effective trapping potential and
prevent the separation of the condensate and the string, thereby allowing
solutions which are more electric (and therefore of a smaller radius) 
than in the original theory.
Below we demonstrate that the effective trapping potential,
and its dependence on $\chi$, can vary with the choice of the parameters
 in the Lagrangian.
In particular, the parameters used in this paper (chosen to allow
explicit exact infinite string solutions carrying current and charge)
produce a trapping potential which does not decrease as $\chi$ increases,
in contrast to some other choices. This suggests that the parameter values used 
have allowed the construction of strongly electric vortons that may not exist
for some other parameter values.

To study the trapping potential we introduce the normalized interaction
\be
\Gamma(\chi)=\frac{\int_{-\infty}^{\infty}\phi^2|\sigma|^2\,dx}
{\int_{-\infty}^{\infty}\eta_\phi^2|\sigma|^2\,dx}
\label{trapping}\ee
which we refer to as the splitting parameter. In this definition
$\phi$ and $\sigma$ are the infinite straight kink and condensate fields
respectively, for a given value of $\chi.$ By construction we have that
 $0<\Gamma<1$
and the larger the value of $\Gamma$ the weaker the trapping potential,
since the condensate samples less of the kink field away from its vacuum value.
This is why we refer to $\Gamma$ as the splitting parameter, since larger
values of $\Gamma$ suggest a greater tendency to instability via splitting.

\begin{figure}
\begin{center}
\includegraphics[width=12cm]{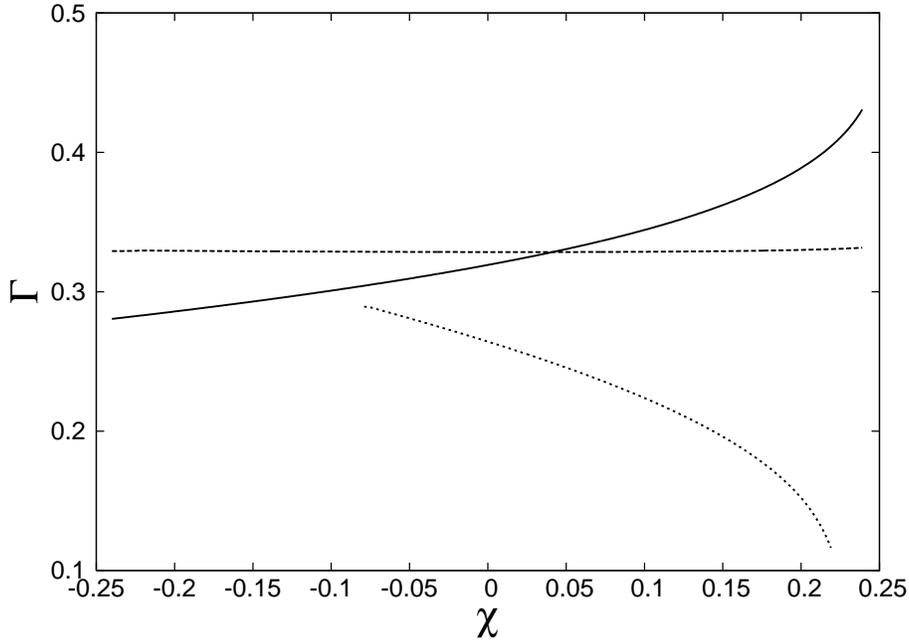}
\caption{The splitting parameter $\Gamma(\chi)$ for three choices
of the parameters  $\eta_\phi, 
\eta_\sigma, \lambda_\phi, \lambda_\sigma, \beta$ :
 solid curve for parameters (\ref{epss});
dashed line for parameters (\ref{param1});
dotted line for parameters (\ref{param3}).
}
\label{fig-gamma}
\end{center}
\end{figure}

For the parameter values (\ref{param1}) used in this paper, the exact
solution can be used to compute $\Gamma(\chi)$ explicitly, yielding 
the result $\Gamma(\chi)=1/3,$ independent of $\chi.$ This is 
a special situation and is not generic. 
As an example, consider the parameter values
used in the study of cosmic vortons in \cite{LS}, namely \be \eta_\phi=1, \quad
\eta_\sigma=\frac{1}{2}, \quad \lambda_\phi=\frac{3}{2}, \quad
\lambda_\sigma=10, \quad \beta=\frac{3}{2}. \label{epss} \ee
For this parameter set there are no exact solutions available and
hence the fields and $\Gamma(\chi)$ must be computed numerically. The result
is displayed as the solid curve in Figure~\ref{fig-gamma} for the allowed
range of $\chi$ (for comparison the dashed line is a numerical
computation for the parameter set (\ref{param1}) and is a test on
the numerical accuracy since the exact result $\Gamma(\chi)=1/3$ is
known in this case). Figure~\ref{fig-gamma} reveals that $\Gamma(\chi)$
increases with $\chi$ for the parameters (\ref{epss}), and suggests
that the splitting instability is more severe for this parameter set,
hence making it more difficult to find strongly electric (and therefore small
radius) vortons. We have been able to compute some kinky vortons with this
parameter set but the solutions indeed appear to exist only in a regime
which is much closer to the chiral limit. This fact, together with
the absence of exact solutions, makes it a more difficult problem
to study in detail.

As mentioned above, one approach to partially overcome the splitting instability
is to modify the power in the interaction term \cite{LS}. However,
the above results suggest that an alternative possibility is to find a more
favourable region of parameter space. The parameter set (\ref{param1})
appears to be a better choice than (\ref{epss}) but there is no reason to
believe that the parameters which allow exact solutions are the best choice 
from this point of view. As a further example, consider the parameter
set
\be \eta_\phi=1, \quad
\eta_\sigma=1, \quad \lambda_\phi=3, \quad
\lambda_\sigma=2, \quad \beta=2 \label{param3}. \ee
The associated splitting parameter $\Gamma(\chi)$ is shown as 
the dotted curve in Figure~\ref{fig-gamma}, for the allowed range
of $\chi.$ In this case $\Gamma(\chi)$ decreases as $\chi$ increases,
suggesting that the parameters (\ref{param3}) could be an improvement
over the values (\ref{param1}) in searching for small electric vortons.
This issue, and particularly its implications for cosmic vortons, is 
currently under investigation.

\section{Conclusion}
In this paper we have introduced and studied kinky vortons, which
are (2+1)-dimensional analogues of cosmic vortons. The lower
dimensional system allows some exact results to be obtained and
greatly simplifies both the analytic and numerical treatment. We
find that there are remarkable similarities between kinky and cosmic
vortons, and that many of the expected properties of cosmic vortons
can be realized in the kinky vorton system.

Our study of kinky vortons has revealed some of the difficulties
that are responsible for the current limited results on cosmic
vortons and has suggested some possible approaches to tackle these
issues. Investigations of cosmic vortons are currently in progress
which make use of the results presented here.

Most studies of cosmic vortons apply a string approximation 
(see \cite{Ca} and references therein) which should
be valid in the thin ring limit, where the vorton radius is much larger
than the cosmic string width. The simulations of kinky vortons suggests
that this will provide a good approximation providing the vorton indeed
exists and is stable. However, the existence and stability 
appear to require a detailed analysis of the full field theory equations,
so some caution may need to be applied when using results derived within
the string approximation. 

It is perhaps worth pointing out that vorton-like objects occur as
classical solutions in several field theories describing condensed
matter systems \cite{BCS,Coo,Su}, so in the future it may be possible to
study these objects experimentally within this setting.

\section*{Acknowledgements}
This work was supported by the PPARC special programme grant
``Classical Lattice Field Theory''.
The parallel computations were performed on 
the Durham University HPC cluster HAMILTON, and
COSMOS at the National Cosmology Supercomputing Centre in Cambridge.\\

\section*{Note Added}
During the final stages of the preparation of this paper an
interesting preprint \cite{RV} appeared on the arXiv  containing
some numerical results on the construction of cosmic vortons.
However, the cosmic vortons in \cite{RV} are in a very different
regime to those discussed in this paper. In particular, the vorton
radius is almost equal to the cosmic string width, so they are far
from the thin ring limit. Indeed, only very low winding numbers,
$N\le 5$, could be computed and it was acknowledged that {\em the
construction of vortons in the thin ring limit remains a numerical
challenge.} The vortons in \cite{RV} were constructed by perturbing
away from the sigma model limit
$\lambda_\phi=\lambda_\sigma\rightarrow\infty$ and
$\beta\rightarrow\infty$ and it was noted that such vorton solutions
are essentially the Skyrmion solutions obtained by the current
authors \cite{BCS} in a related model of a two-component
Bose-Einstein condensate.

\section*{Appendix A}\news\renewcommand{\theequation}{A.\arabic{equation}}

In this appendix we provide some formulae for kinky vortons for general values of the parameters $\lambda$, $\eta_{\phi}$ and
 $\eta_{\sigma}$, while maintaining the condition $\lambda=\lambda_\phi=\lambda_\sigma=2\beta$. If we define
 $\alpha=(\eta_\sigma/\eta_\phi)^2$ and ${\hat\chi}=\chi/{m}_{\phi}^2\,$ then the constraint (\ref{cond5})
becomes
\be
{1\over 2}-\alpha<{\hat\chi}<1-\alpha\,,
\ee
and the solutions (\ref{gstring}) can be written as
\begin{eqnarray}
\phi&=&\eta_\phi\tanh\left[{m}_\phi x\sqrt{1-\alpha-{\hat\chi}}\right]\,,\cr |\sigma|
 &=&\sqrt{2}\eta_{\phi}\sqrt{\alpha-{1\over 2}+{\hat\chi}}\,{\rm sech}\left[{m}_\phi x\sqrt{1-\alpha-{\hat\chi}}\right]\,.
\end{eqnarray}
Using these formulae one can show that 
\be
Q={16\pi R\omega\over\lambda}{(\alpha-\frac{1}{2}+{\hat\chi})\over\sqrt{1-\alpha-{\hat\chi}}}\,,
\ee
and $\hat\chi$ satisfies the cubic
$$
16{\hat\chi}^3+\left[32\left(\alpha-\frac{1}{2}\right)+\frac{16N^2}{{\hat R}^2}\right]{\hat\chi}^2
+\left[16\left(\alpha-\frac{1}{2}\right)^2+32\left(\alpha-\frac{1}{2}\right)\frac{N^2}{{\hat R}^2}+
\left(\frac{Q\lambda}{4\pi{\hat R}}\right)^2\right]{\hat\chi}
$$
\be
+\left[16\left(\alpha-\frac{1}{2}\right)^2 N^2 +(\alpha-1)\left(\frac{Q\lambda}{4\pi}\right)^2\right]\frac{1}{{\hat R}^2}=0
\ee
where $\hat R={m}_\phi R$. The energy is then given by
\be
E={2\pi\eta_{\phi}^2\over\sqrt{1-\alpha-{\hat\chi}}}\bigg(\frac{{\hat R}}{3}\bigg[
2(2\alpha+1)(2-2\alpha-{\hat\chi})+{4}{\hat\chi}^2\bigg]
 +{1\over\hat R}\bigg[4N^2\left(\alpha-\frac{1}{2}+{\hat\chi}\right)+{1-\alpha-{\hat\chi}\over \alpha-
\frac{1}{2}+{\hat\chi}}\left({Q\lambda\over 8\pi}\right)^2\bigg]\bigg)\,.
\ee
For the values $\eta_\phi=1\,$, $\lambda=2$ and $\alpha=3/4$, as used in the rest of this paper, these expressions
reproduce those presented earlier.

\section*{Appendix B}\news\renewcommand{\theequation}{B.\arabic{equation}}
In this appendix we provide an analysis to estimate the value of $\chi$ below
which a pinching instability is expected to exist. This is a lower dimensional
analogue of a similar calculation performed for cosmic vortons in \cite{LS}.

The instability is studied for a straight infinite kink string in the magnetic
 regime. Using the Lorentz invariance of the theory we can restrict to
the case $\omega=0$ and therefore $\chi=-k^2.$ The task is to consider
the stability of the kink string fields (\ref{string})
 \be
\phi_0=\tanh\bigg(\frac{x\sqrt{1-4\chi}}{2}\bigg), \quad\quad
\sigma_0=e^{iky}\sqrt{\frac{1+4\chi}{2}}{\rm sech}\,
\bigg(\frac{x\sqrt{1-4\chi}}{2}\bigg).
\label{astring}
\ee
An exact analysis of the stability of this solution would require
a numerical approach, so to make analytic progress we consider
a perturbation of the condensate field of the form
\be
\sigma=\sigma_0+|\sigma_0|\nu, \label{pert}
\ee
where $\nu(y,t)$ is a complex function of $y$ and $t$ but is
independent of $x.$ There are no exact negative modes with the
above specified $x$ dependence, as can be seen from the linearized
equation
\be
|\sigma_0|(\partial_t^2\nu-\partial_y^2\nu)-\nu|\sigma_0|''+\nu|\sigma_0|
(2|\sigma_0|^2-\frac{3}{4}+\phi_0^2)+\bar\nu|\sigma_0|\sigma_0^2=0.
\ee
However, we can consider integrating this equation over the string cross-section
by multiplying by $|\sigma_0|$ and integrating over $x.$ 
This produces the equation
\be
\Sigma_2(\partial_t^2\nu-\partial_y^2\nu)+\nu(\chi\Sigma_2+\Sigma_4)
+\bar\nu\Sigma_4e^{2iky}=0,
\label{av}\ee
and we can now search for negative modes of this equation of the
form 
\be
\nu(y,t)=e^{\Lambda t}e^{iky}u(y),
\label{defy}
\ee
where $u(y)$ is a complex function of $y$ and we require that $\Lambda^2>0$
so that $\Lambda$ is real with a positive value corresponding to the
required negative mode.
The eigenvalue equation for $u(y),$ obtained by substituting (\ref{defy})
into (\ref{av}), is
\be -u''-2iku'+(u+\bar u)\Sigma+\Lambda^2u=0,
\label{ueqn}
\ee 
where we have defined $\Sigma=\Sigma_4/\Sigma_2.$
Note that an equivalent method to derive this equation is to substitute
the perturbation (\ref{pert}) into the expression for the energy and
integrate over the string cross-section, then require negative modes
for the resulting effective energy.
 
Splitting $u$ into real and imaginary parts as $u=u_1+iu_2$ gives
the coupled equations
\be
-u_1''+2ku_2'+(2\Sigma+\Lambda^2)u_1=0, \quad\quad
-u_2''-2ku_1'+\Lambda^2u_2=0.
\ee
Expanding in terms of fourier modes 
$u_1=a_1\cos(py)$ and $u_2=a_2\sin(py)$ yields the equation
\be
\pmatrix{p^2+2\Sigma+\Lambda^2 & 2kp\cr 2kp & p^2+\Lambda^2}
\pmatrix{a_1\cr a_2}=0.
\ee
A non-zero solution exists only if the determinant vanishes,
which gives the condition
\be\Lambda^2=-(p^2+\Sigma)\pm\sqrt{\Sigma^2-4\chi p^2}.\ee
Requiring a solution with $\Lambda^2>0$ produces the constraint
\be
p^2+2\Sigma+4\chi<0,
\ee
and hence 
\be
\chi<-\frac{\Sigma_4}{2\Sigma_2}.
\ee
Using the exact expressions (\ref{sigma24}) this constraint
is equivalent to $\chi<-\frac{1}{10}.$

The above analysis has neglected the curvature of the vorton
and also integrated over the string cross-section, rather than
solving the full linearized equations, but it is expected to
give a reasonable estimate of the critical value of
$\chi$ below which the magnetic instability appears.

\end{document}